\documentclass[prl,preprint]{revtex4}
\usepackage{amsmath}
\usepackage[dvips]{graphicx}
\usepackage{bm}

\newcommand{\Eref}[1]{Eq.~(\ref{#1})}

 \newcommand{\NOCR}{Titov:85Dism,Titov:96b,Petrov:02}

\begin{document}
\title{On search for nuclear Schiff moment in liquid xenon}
\author{T.A.\ Isaev}
\email{timisaev@pnpi.spb.ru}
\affiliation{Petersburg Nuclear Physics Institute, Gatchina, 188300, Russia}
\author{A.N.\ Petrov}
\affiliation{Petersburg Nuclear Physics Institute, Gatchina, 188300, Russia}
\author{N.S.\ Mosyagin}
\affiliation{Petersburg Nuclear Physics Institute, Gatchina, 188300, Russia}
\author{A.V.\ Titov}
\email{titov@AT1770.spb.edu} \affiliation{Chemistry Department,
St.-Petersburg State University, St.-Petersburg, 198504, Russia}
\affiliation{Petersburg Nuclear Physics Institute, Gatchina,
188300, Russia}

\begin{abstract}

 A parameter of the P,T-odd Hamiltonian characterizing interaction of the
 nuclear Schiff moment with the gradient of electronic density on the Xe
 nucleus is calculated for an isolated Xe atom and for liquid xenon.  We
 use more realistic model of liquid medium than the spherical cell model
 used in [B.Ravaine and A.Derevianko Phys.\ Rev.\ A {\bf 69},
 050101(R)(2004)]. Qualitatively different results for enhancement of
 the P,T-odd effect in liquid xenon are obtained when polarization of the
 medium is taken into account. Thus, proper choice of the liquid phase
 model is crucially important even for calculation of the properties
 dependent mostly on the electronic density near a nucleus.
\end{abstract}

\maketitle

\section{Introduction.}

 Schiff moment is a space parity (P) and time invariance (T) violating
 (P,T-odd) nuclear moment that is expected to exist because of P,T-odd
 interactions between nucleons and/or existence of uncompensated contribution
 from the permanent electric dipole moment (EDM) of the unpaired nucleon,
 enhanced (or, in principle, suppressed) by nucleus core polarization
 \cite{Ginges:04}. The conventional choice for experiments on search for the
 Schiff moment up to the present time has been heavy atoms or polar diatomic
 molecules containing a heavy atom because of great enhancement of the P,T-odd
 effects in such systems.  Recently some new approaches to search for the
 Schiff moment were suggested. Their principal feature is using more complex
 systems, particularly, solid state \cite{Mukhamedjanov:04} and liquid xenon
 (LXe) \cite{Romalis:01b, Sobelman:05} have been considered for the
 corresponding experiments. Though solids and liquids have a number of
 advantages first of all because of
 the higher statistical factor in comparison with that for the atomic/molecular
 beam/cell experiments, reliable calculations of their electronic structure
 (always required in the experiments for the interpretation of the data in
 terms of constants of the P,T-violating weak interaction) are, certainly,
 substantially more complicated than for the diatomic molecules.  Even in a
 closed-shell Van der Waals system such as liquid xenon the choice of the
 appropriate model for the medium is not trivial for the purpose of reliable
 theoretical study.  For example, in \cite{Ravaine:04} the cell model of liquid
 xenon was used. In that model the single Xe atom is confined in the
 spherically symmetric cavity and a density-dependence factor is introduced
 through variation of the cavity radius. Using the cell model could be
 justified by the fact that LXe is non-polar liquid with the dielectric
 constant close to unity. In \cite{Ravaine:04} authors found that the effect of
 P,T-violation is suppressed in their model for liquid xenon on about $40\%$ in
 comparison with an isolated (free) atom.  However, the influence of
 polarization of the medium on the P,T-odd property
 was not studied. The purpose of the current work is calculation of the P,T-odd
 effect caused by the Schiff moment of the $^{129}$Xe nucleus
 using more realistic models of LXe than the cell model and studying
 the contribution from polarization of the
 medium on the P,T-odd property in liquid xenon.

\section{Methods and approaches.}
\label{Methods}

 Electronic structure calculations of the isolated Xe atom and
 liquid medium are carried out using two-step method (see \cite{Titov:06amin}
 and references therein), the most principal features of which are:
\begin{enumerate}
  \item A two-component electronic (pseudo)wave function is first obtained in
  calculation with the generalized relativistic effective core potential
  (GRECP) for Xe with either 8 or 26 explicitly treated electrons, providing
  proper electronic density in the valence and outer core regions.
  \item The proper shape of the four-component molecular spinors in the inner
  core region of the Xe atom is restored using the non-variational one-center
  restoration (NOCR) scheme \cite{\NOCR} followed by calculations of P,T-odd
  parameters essentially depending on electronic density near the heavy
  nucleus.
\end{enumerate}

 In the present work, the finite field method \cite{Kunik:71} is used for the
 required property calculations in the framework of the relativistic coupled
 cluster method with single and double cluster amplitudes (RCC-SD)
 \cite{Eliav:96, Kaldor:04a}.  The latter was chosen because it is known to
 work well for the closed shell systems.

 Generalized correlation-consistent (GC) spin-orbital atomic basis sets
 \cite{Mosyagin:00,Isaev:00} are used in xenon electronic structure
 calculations. GC basis sets were specially optimized for calculation of the
 ``core-type'' properties and, besides, were augmented with the polarization
 $d$-function optimized with respect to the value of atomic polarizability
 $\alpha_p$. As a result, basis set [4s,6p,4d,2f] was constructed for
 calculations with 8-electron GRECP and [5s,6p,4d,2f] for 26-electron GRECP.
 All molecular spinors are then restored as one-center expansions on the
 $^{129}$Xe nucleus.  The nucleus is modelled as a uniform charge distribution
 within a sphere
 with the radius $r_{\rm nucl} = 6.2~{\rm fm}\equiv 1.17\times10^{-4}$ a.u.
 The gaussian expansion of basis sets and results of atomic calculations can be
 found elsewhere (see http://qchem.pnpi.spb.ru/Basis/Xe). To calculate the
 atomic polarizability and parameter of the P,T-odd Hamiltonian, $X$ (see
 \Eref{X}), we have applied electric field created by four point electric
 charges located on $z$ axis. In order to reach sufficient homogeneity of
 electric field at the xenon cluster the charges \{-80,~5,~-5,~80\} a.u. were
 located in points \{-100,~-50,~50,~100\} (a.u., on $z$ axis), thus electric
 field 0.012 a.u.\
 was created at the coordinate source. Below we will use the atomic units
 unless other is stated explicitly. Such electric field is weak enough to
 neglect higher-order terms in calculations of $\alpha_p$ and parameter $X$,
 but sufficiently strong to avoid problems with numerical accuracy.  The field
 homogeneity and strength were checked in the atomic SCF and RCC-SD
 calculations of $\alpha_p$ with homogenous field and $X$ taking the field
 0.0012 (ten times weaker than the originally used one).

 The parameter $X$ of the P,T-odd Hamiltonian \cite{Petrov:02}
 was calculated for the Xe atom in the presence of electric field:
\begin{equation}
   X=\frac{2\pi}{3} \left[
     \frac{\partial}{\partial z}\rho_{\psi}(\vec{r})
      \right] _{x,y,z=0}\ ,
  \label{X}
\end{equation}
 where $\rho_{\psi}(\vec{r})$ is the electronic density calculated from the
 wave function $\psi$.  Such form of $X$ is derived from contact form of the
 P,T-odd interaction of the Schiff moment with the electronic density. In
 \cite{Ginges:04} more sophisticated form of P,T-odd interaction was suggested,
 where the ``finite-volume'' Schiff moment potential is used.  In the atomic
 Xe calculations it was found that the ``finite-volume'' form of Schiff
 interaction does not lead to noticeable changes in the results of the P,T-odd
 properties calculation in comparison with the contact one \cite{Dzuba:06priv}.
 Connection of the parameter $X$ with the parameters typically evaluated in
 atomic calculations of the EDM induced by the Schiff moment can be derived
 following \cite{Dzuba:02}:
$$
   X=\frac{1}{6} \frac{\bm{S} \cdot \bm{n}}{ \bm{D_e}\cdot \bm{E}},
$$
 where $\bm{S}$ is the {S}chiff moment of $^{129}$Xe nucleus, $\bm{n}$ is a
 unit vector along the molecular axis (which in our case coincides with the
 direction of external electric field), $\bm{D_e}$ is the atomic EDM induced by
 the nucleus {S}chiff moment, $\bm{E}$ is an external electric field applied to
 the Xe atom.

\paragraph*{Atomic calculation.}

 To analyze the different correlation contribution we have performed
 calculations of $X$ and $\alpha_p$ for an isolated Xe atom with the different
 numbers of correlated electrons and the level of correlation treatment. The
 results are summarized in Table \ref{Xe_at}. Spin-averaged GRECP (AGREP) was
 used in SCF calculations, while the spin-orbit GRECP component was also
 included in RCC-S and RCC-SD calculations. The difference between 26-electron
 GRECP/RCC-S (which is roughly analogous to SCF calculation accounting for the
 spin-orbit (SO) interaction in the closed-shell case) and all-electron DHF
 results is within 10\%, which can be connected with the incompleteness of
 basis sets and polarization contributions from inner shells. As the main
 purpose of the current work is not calculating the P,T-odd effects with very
 high precision but rather investigating the influence of the liquid medium on
 the P,T-odd effect(s) we use 8-electron GRECP in subsequent calculations to
 reduce computational expenses.  It is seen from Table \ref{Xe_at} that
 contribution to the $X$ value from the SO effects is only about 5\%,
 thus we are not calculating further RCC-S values separately and rather going
 from AGREP/SCF right to GRECP/RCC-SD values accounting for both correlation
 and SO-effects.
 Note, however,
 that contribution of the SO interaction taken into account at the GRECP/RCC-SD
 or GRECP/RCC-S calculation (for the ``outer parts'' of explicitly treated
 shells) is different from the SO contributions taken into account at the
 AGREP/SCF stage (for the inner shells excluded from the GRECP calculations
 since they were treated as SO-split atomic spinors at the GRECP generation)
 and at the NOCR stage (for the ``inner parts'' of the shells explicitly
 treated within GRECP calculations).

\squeezetable
\begin{table*}[!]
\caption
 {Calculated parameter $X$ and scalar polarizability $\alpha_p$ for the
 $^{129}$Xe single atom, compared with values from \protect\cite{Dzuba:02}.
 Relativistic coupled cluster methods with single (RCC-S) and single and double
 cluster amplitudes (RCC-SD) are used with 8- and 26-electron Xe GRECPs. All
 values are in a.u.}
\begin{tabular}{lccccccccccc}
& \multicolumn{2}{c}{8 electrons}& & \multicolumn{3}{c}{26 electrons}& &
\multicolumn{2}{c}{All-electron$^a$}\\
          \hline
          & AGREP/SCF & GRECP/RCC-SD& &AGREP/SCF &GRECP/RCC-S& GRECP/RCC-SD& & DHF & TDHF &
          \multicolumn{2}{c}{Experimental}\\
          \hline
$X$        &     238  &  213  & &   186  &   176  & 173   &  & 165 & 213   &  &  \\
$\alpha_p$ &    26.7  &       & &   26.6 &   26.6 & 27.2  &  & 26.9 & 27.0 & 27.16$^b$ & 27.815$^c$ \\
\hline
\end{tabular}
\\
\vspace{0.5cm}
\begin{flushleft}
 \noindent $^a$ Four component atomic calculations
           by Dirac-Hartree-Fock method (DHF) and time-dependent DHF (TDHF).\\
 \noindent $^b$ Reference \cite{Kumar:85}. \\
 \noindent $^c$ Reference \cite{Hohm:90}.
\end{flushleft}

 \label{Xe_at}
\end{table*}

\paragraph*{Liquid xenon model.}

\begin{figure}
\includegraphics[bb = -100 0 500 300, scale=0.8]{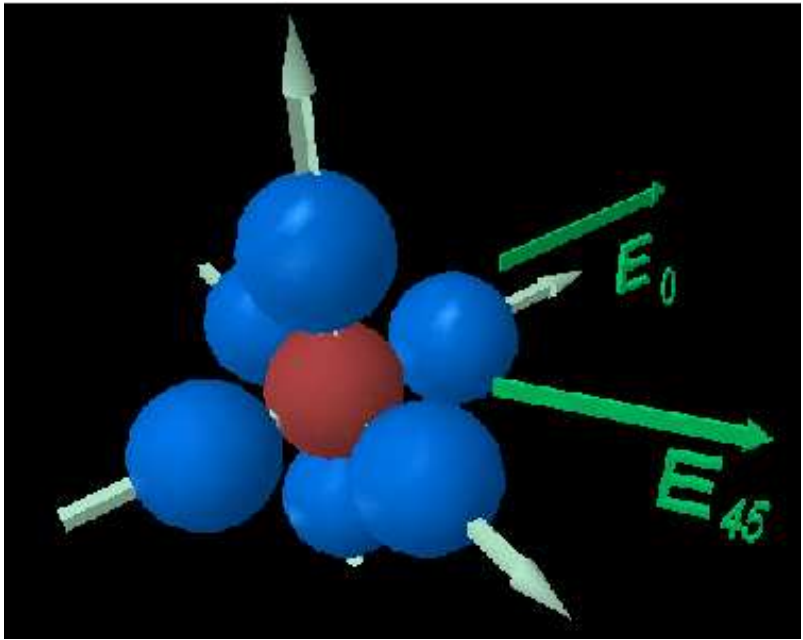}
\caption{Series $\bm{I}$ lattice cell geometry and electric field. In the 1st
  geometry electric field E$_0$ is directed along $z$ axis. In the 2nd geometry
  electric filed E$_{45}$ is directed under $45^o$ to $z$ axis.}
 \label{Xe_a}
\end{figure}

\begin{figure}
\includegraphics[bb = -100 0 500 300, scale=0.8]{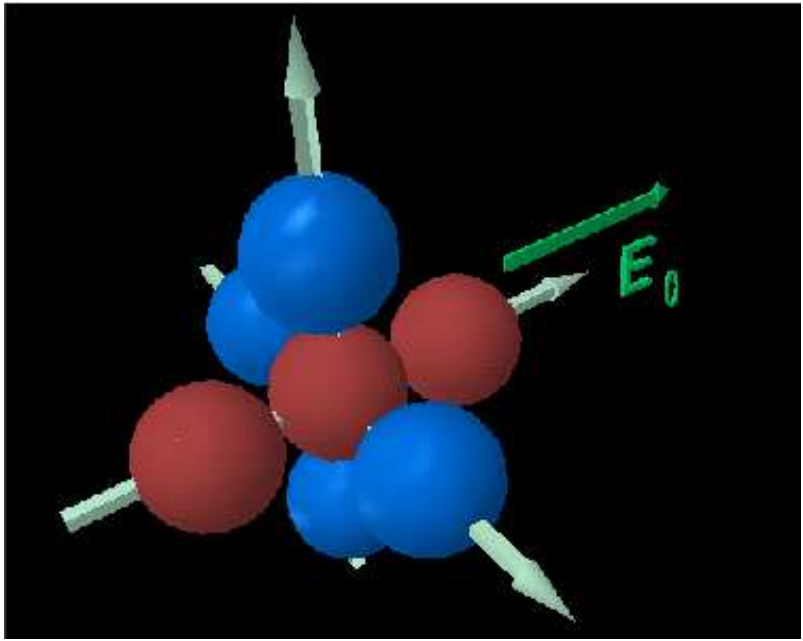}
\caption{Series $\bm{II}$ lattice cell geometry and electric field.}
 \label{Xe_b}
\end{figure}

 In liquid xenon calculations we used the simple cubic crystalline lattice (see
 Fig.~\ref{Xe_a}).  It is known (e.g., see \cite{Stampfli:91}), that the
 decrease of the density in liquid xenon as compared to the solid state phase
 is mainly due to decreasing the average number of the neighboring atoms (i.e.\
 introducing vacancies) according to the ratio
 \mbox{$\rho_{\rm liq}$/$\rho_{\rm sol}$}, where $\rho_{\rm liq}$
 $(\rho_{\rm sol})$ is the density of
 liquid (solid) xenon.  On the other hand, at atmospheric pressure
  \mbox{$\rho_{\rm liq}$/$\rho_{\rm sol}$}$\approx$0.83,
 thus, the lattice model can be used for the liquid phase description without
 introducing principal error.  Though in solid xenon the crystalline lattice is
 the face-centered cubic one (FCC), there are some reasons to use simple cubic
 (SC) lattice model.
%
%
 In such lattice the number of neighboring atoms is reduced twice in comparison
 with the FCC lattice, nevertheless, the minimal number of the atoms which
 required for adequate analysis of the effects of the liquid medium is kept.
 As it will be seen from the further discussion, the atoms which are
 in the plane perpendicular to the electric field and going through the central
 Xe atom practically do not influence the value of
 $X$.  Thus, on the one hand, high-symmetry SC lattice allows us to
 analyze the effect of the liquid medium with minimal computational efforts ---
 the only six neigbours reproduce the volume effect reasonably well whereas
 the number of explicitly treated electrons is reduced almost twice,
 and, on the other hand, it reflects the most principal features of
 the liquid medium.  Besides, high symmetry of the SC lattice simplifies the
 analysis of different density- and polarization-dependence effects keeping
 reasonable computational expenses within the $C_{2v}$ symmetry implemented in
 the code used by us.
 Electric field modelled by four point charges described in section ``Methods
 and approaches'' was applied to the cluster of seven xenon atoms. We used two
 geometries of relative arrangement of the Xe atoms and electric field, both
 geometries are shown on Fig.~\ref{Xe_a}.  The other geometries were not
 considered because they have the symmetry which is not implemented in our
 codes or require to use lower symmetry that leads to high computational
 expenses.  Blue color means that all explicitly taken eight electrons of the
 given Xe atom were frozen by the level-shift technique \cite{Bonifacic:74}
 after the atomic spin-free AGREP/SCF calculations, while red color means
 involving all eight electrons to SCF and RCC procedure. Two series of
 calculations were performed denoted below as $\bm{I}$ and $\bm{II}$. In series
 $\bm{I}$ all electrons of all atoms but the central Xe atom were frozen after
 the atomic AGREP/SCF calculations (Fig. \ref{Xe_a}) In series $\bm{II}$
 electrons in the 1st geometry were frozen only for atoms which are not on axis
 $z$ (Fig.~\ref{Xe_b}).  The interatomic distance $R_l$ in the lattice cell
 (lattice constant) was changed from 6.0 a.u.\ to 20 a.u. The value of
 $R_l=6.6$ corresponds to the elementary {\it lattice} cell of the {\it same
 volume} that was taken in \cite{Ravaine:04} for the {\it spherical} cell.
 The more realistic value for the equilibrium lattice constant $R_l$ is
 connected with the parameters of the FCC lattice for the solid xenon density
 (in this case the distance between the closest neighbors in the FCC lattice is
 $R_{l-FCC}=8.6$). As explaned above, we, however, use the model of simple
 cubic lattice with $R_l=n^{-1/3}=8.0$, where $n$ is the density of liquid
 xenon (for the solid xenon density the corresponding constant would be
 $R_l=7.5$).  In any case, as we will see further, the reasonable choice of
 such or another $R_l$ is not principal for our conclusions.


\section{Results and discussion}
 \paragraph[]{Series $\bm{I}$:}

\begin{figure}
\includegraphics[bb = -100 0 400 300, scale=1.0]{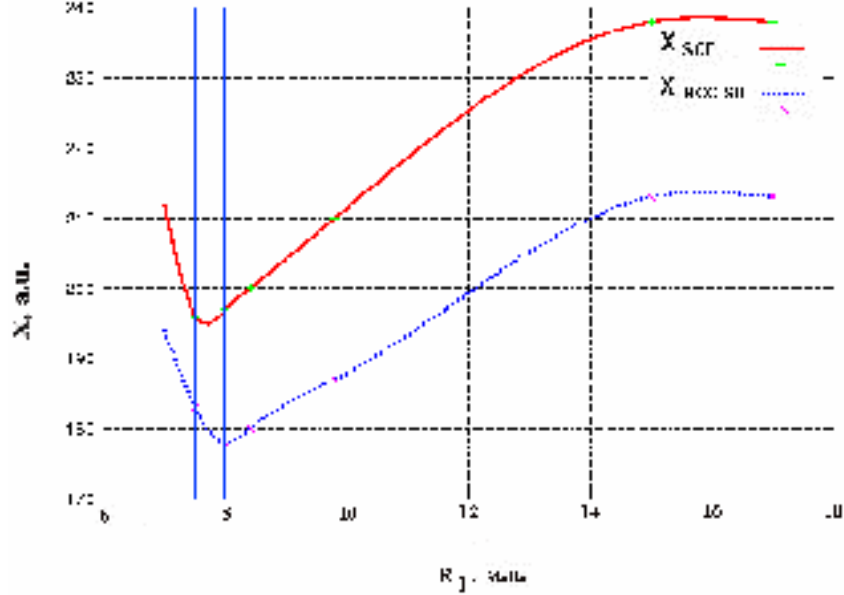}
\caption{Series $\bm{I}$ $X$ value as a function of the lattice cell
  size. $X_{\rm SCF}$ curve corresponds to AGREP/SCF result,
  $X_{\rm RCC-SD}$ stands for GRECP/RCC-SD results.
  Vertical lines are the functions $R_l=7.5$ and $R_l=8.0$,
  see section ``Methods and approaches''.
 }
 \label{ser_a_fig}
\end{figure}

 The results of AGREP/SCF and GRECP/RCC-SD calculations are presented on
 Fig.~\ref{ser_a_fig}.  Qualitative agreement with the results of
 \cite{Ravaine:04} is seen from that picture, though the suppression of P,T-odd
 effect in liquid phase is only about 20\%, in contrast to 40\% obtained in
 \cite{Ravaine:04}.  An interesting peculiarity which is seen on
 Fig.~\ref{ser_a_fig} is the increase of the $X$ value when $R_l\leq7.5$. The
 natural reason for such behavior can be the increase of the amplitude of the
 wavefunction on the Xe nucleus due to compression of the tail of the
 electronic density in the valence region.  For $R_l>7.5$ the wavefunction
 amplitude increasing on the nucleus is less important than effect of
 suppression of the polarizability of the Xe atom in the ``frozen'' medium.
 Actually, the same is observed in the spherical cavity model (where
 polarizability of the valence shells is suppressed even stronger) that results
 in decreasing the value of the P,T-odd effect. We have seen very weak
 dependance (within 1 \%) of $X$ value from either 1st or 2nd geometry used. In
 further calculations we used only the 1st geometry, such choice for series
 $\bm{II}$ is explained below.

 \paragraph[]{Series $\bm{II}$:}
\label{ser_b}

\begin{figure}
 \includegraphics[bb = -100 0 800 900, scale=0.6]{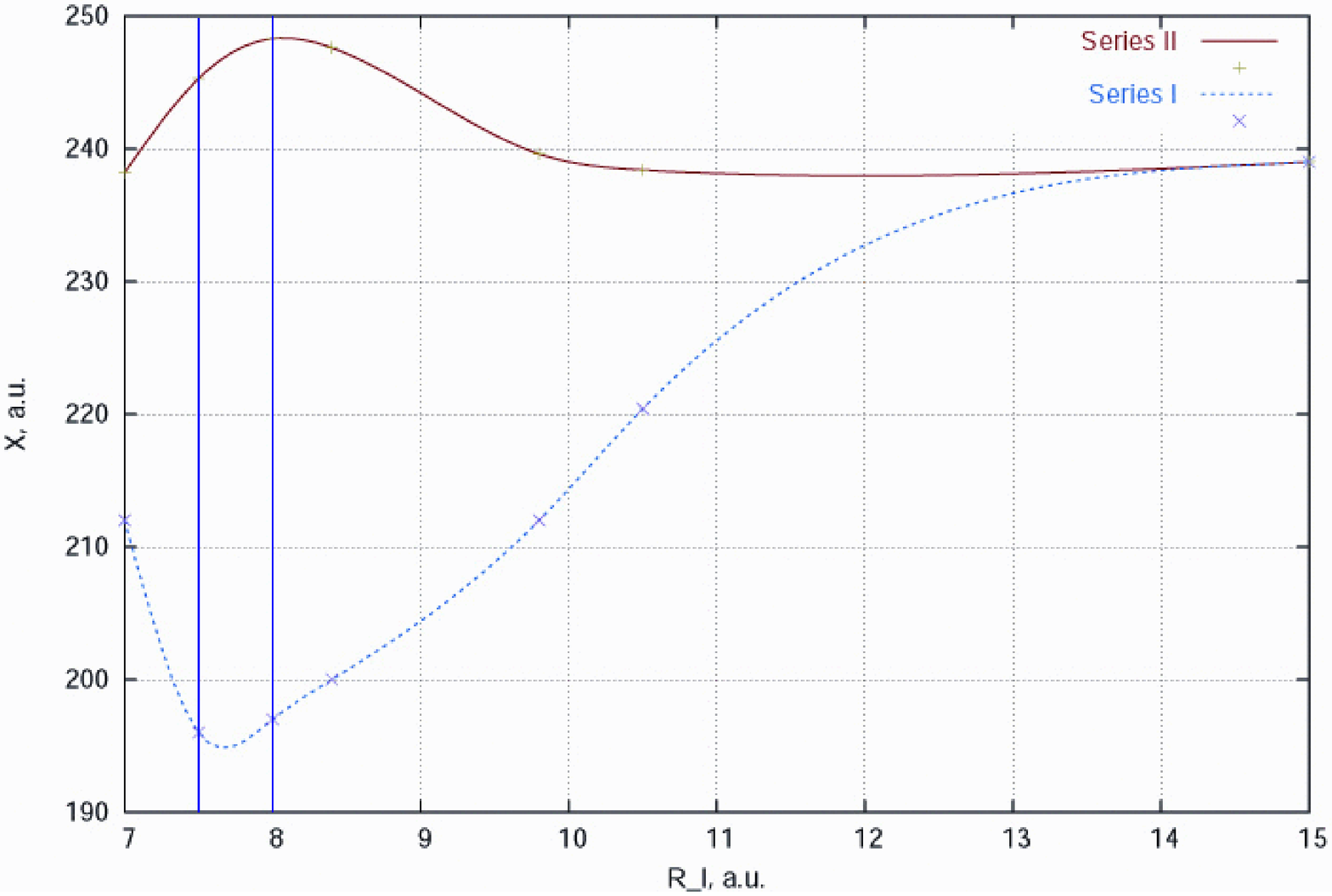}
\caption{Series $\bm{I}$ and $\bm{II}$ AGREP/SCF $X$ values as a function of
         the cell size.  Vertical lines are the functions $R_l=7.5$ and
         $R_l=8.0$, see section ``Methods and approaches''.
 }
 \label{ser_b_fig}
\end{figure}

 The results of AGREP/SCF and GRECP/RCC-SD
 calculations are presented on
 Fig.~\ref{ser_b_fig}.  The test calculations were performed first with
 $R_l=7.5$ to estimate influence of the neighboring atoms. It turned out that
 accounting for polarization/correlation effects for the atoms which are not on
 $z$ axis (that is directed along the electric field) practically does not
 change the value of $X$, that looks rather natural.  Thus, in subsequent
 calculations we accounted only for polarization/correlation of the electrons
 belonging to the atoms on $z$ axis.  Reduced GC basis sets described in
 section ``Methods and approaches'' were used --- on the central Xe atom GC
 basis was [4s,6p,4d] and on two neighbor atoms basis sets were reduced to
 [4s,6p]. Though, formally, $\{s,p\}$ basis set is not good for description of
 polarizaton/correlation effects on the atom, functions from the other atom are
 partially capable to account for those effects.

 One can see that accounting for polarization of the neighboring atoms leads to
 crucial change in the enhancement of the Schiff moment in liquid xenon. The
 P,T-odd effect is practically {\it not suppressed} but rather {\it enhanced}
 in liquid xenon.

 The physical reason for such behavior is interference of the polarization of
 the liquid medium and the density-dependence effects (compression of the
 electronic density in the valence region).  The following basic
 interpretations are the most reasonable to consider for the polarization
 contribution:
\begin{enumerate}
 \item
 the tails of electronic density from {\it neighboring} atoms penetrate to the
 inner core of the central Xe atom.  Due to polarization effects the
 contributions to the gradient of the electronic density on the central Xe
 nucleus (which $X$ is proportional to, see \Eref{X}) from neighbors do not
 compensate each other;
 \item
 {\it own} electronic shells of the central Xe atom interact with the outermost
 electronic shells of neighboring atoms, thus affecting the $X$ value. As outer
 shells of neighboring atoms are polarized in the electric field, their
 polarization influences the polarization of the outer shells of the central Xe
 atom.
 Also, the amplitude of the electronic wavefunction could be increasing
 at the $^{129}$Xe nucleus due to compression of the electron wavefunction
 tails in the valence region whereas the polarization of the isolated atom and
 the atom in liquid medium could be comparable.  Such an effect of the ``volume
 compression'' would manifest itself in $1/R_l^3$\,--\,like dependance of the
 $X$ value from $R_l$
 (with $R_l$ for which outermost shells of the neibour Xe atoms start to
 overlap) on Fig.~\ref{ser_b_fig}, but, as our analysis show, the dependance
 X($R_l$), ($9.0<R_l<10$), is linear.  Nevertheless, some combination of the
 above effects is rather taking place.

%
\end {enumerate}
 Though the above discussed effects are undistinguishable physically (at least
 for the case of the xenon cluster), from computational point of view they
 are treated as different ones.
 The spherical cell model in \cite{Ravaine:04} does not account for the above
 effects properly.

\begin{figure}
\includegraphics[bb = 0 0 400 400, scale=0.6]{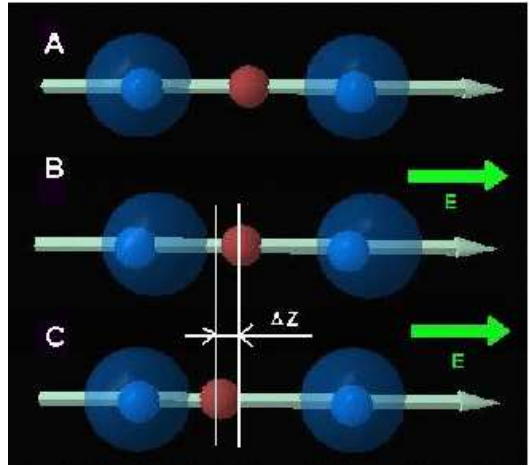}
\caption{Simulation of the influence of electric
  filed through lattice geometry. Only atoms on $z$ axis are plotted. Red color
  designates the nuclei of the central Xe
  atom (its shells explicitly treated in the calculations are not shown here),
  outer shells of the neighboring
  atoms are half-transparent blue, inner shells and nuclei are blue. On picture
  {\bf A} atoms are without electric field, when center of charge of the outer
  electronic density coincides with that of the inner shells and nuclei. On
  picture {\bf B} in the electric field (designated $E$) the centers of charge
  of the outer shells are shifted in respect to inner part. On picture {\bf C}
  outer shells electronic density of the neighboring atoms is shifted so as to
  superpose their charge center to those of neighboring atoms but keep the
  relative position to the central Xe atom the same as in case {\bf R}.}
 \label{Delta_Z}
\end{figure}

 To clarify the situation we carried out two series of calculations.
 In the first one we placed three Xe atoms on $z$ axis and the distance between
 the Xe nuclei was chosen to be $7.5$. The electrons of the central atom were
 frozen, while the electrons of its two neighbors were allowed to relax
 (polarize) in the external electric field, which strength was $0.012$. Thus,
 the only contribution to $X$ came from the electronic density from the
 neighboring atoms, as frozen atomic shells of the central atom had definite
 parity. We obtained $X=1.4$.  On the other hand, the contribution from every
 neighbor was about $X_{s}\approx \pm 24)$ (sign depends on relative position
 with respect to the central atom), thus the great compensation of the
 contributions from the neighboring atoms takes place.
 In the second series we performed calculation of $X$ value almost in the same
 geometry as in series {$\bm I$}.  Electrons of neighboring atoms were frozen,
 while in the central atom eight electrons were correlated. Then the central
 atom was shifted with respect to the coordinate source on the value $\Delta
 z$, see Fig.~\ref{Delta_Z} (picture $\bm{S}$).

\begin{figure}
\includegraphics[bb = 0 0 500 800, scale=0.5]{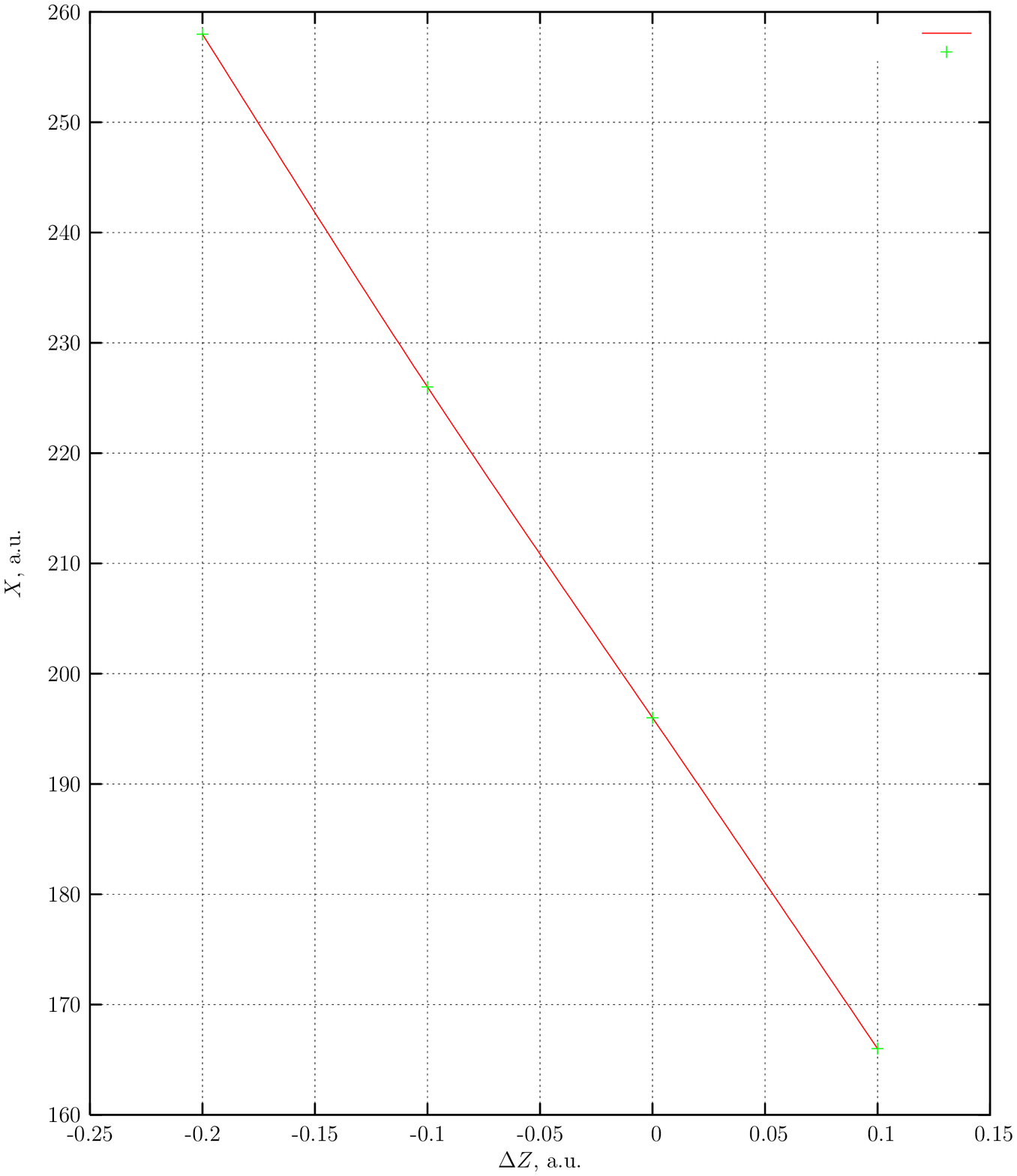}
\caption{AGREP/SCF $X$ value as a function of the shift of the central atom.}
 \label{X_z}
\end{figure}

 As one can see from Fig.~\ref{Delta_Z} such geometry simulates polarization of
 the shells of the neighbor atoms with electric field. The obtained dependance
 of the $X$ value from $\Delta Z$ is given on Fig.~\ref{X_z}.  To get $X$ as
 large as in series $\bm{II}$ one has to take $\Delta Z \approx -0.15$. Taking
 into account atomic Xe polarizability $\alpha_p$ (see Table \ref{Xe_at}) one
 can estimate that $\Delta Z \approx -0.15$ is compatible with
 possible value for the shift of the center of charge density of outer
 $p$-electrons in the field 0.012. Thus dramatic change of the $X$ value in the
 lattice model in comparison with the cell model can be rather attributed to
 the effect of rearrangement of electronic density of the liquid medium.
%

\paragraph*{Acknowledgments.}

 This work is supported by the RFBR grant 06--03--33060.  T.I.\ and A.P.\ are
 grateful for the grant of Russian Science Support Foundation.
 A.P.\ is also supported by grant of Gubernator of Leningrad district.

\bibliographystyle{../bib/apsrev}
\bibliography{../BIBs/JournAbbr,../BIBs/TitovLib,../BIBs/Titov,../BIBs/Kaldor,../BIBs/IsaevLib}
\end{document}